\documentclass{INTERSPEECH2023}

\usepackage{amsmath,graphicx}
\usepackage{cite}
\usepackage{algorithmic}
\usepackage{array}
\usepackage{stfloats}
\usepackage{xcolor}
\usepackage{epsfig,setspace}
\usepackage{float,algorithm}
\usepackage{bm}
\usepackage{multirow, hhline}
\usepackage{makecell}
\usepackage{amssymb}
\usepackage{pifont}
\usepackage{arydshln}
\usepackage{booktabs}
\newcommand{\cmark}{\ding{51}}
\newcommand{\xmark}{\ding{55}}
\newcommand{\tabincell}[2]{\begin{tabular}{@{}#1@{}}#2\end{tabular}}
\usepackage{subfig}
\usepackage{caption}
\usepackage{amsmath}

\usepackage[section]{placeins}

\interspeechcameraready

\title{Use of Speech Impairment Severity for Dysarthric Speech Recognition}
\name{Mengzhe Geng, Zengrui Jin, Tianzi Wang, Shujie Hu, Jiajun Deng, Mingyu Cui, Guinan Li, Jianwei Yu, Xurong Xie, Xunying Liu}

\begin{document}
\bstctlcite{IEEEexample:BSTcontrol}
% \ninept

\maketitle
 
\begin{abstract}
% the following is from professor 
% \textcolor{red}{A key challenge in dysarthric speech recognition is the speaker level diversity attributed to both normal speech factors associated speaker-identity such as gender, and speech impairment severity. The majority of prior researches on addressing this issue focused on using speaker-identity only. To this end, this paper proposes a novel set of techniques to use both speech impairment severity and speaker-identity in dysarthric speech recognition: a) multitask training incorporating severity prediction error; b) speaker-severity aware auxiliary feature adaptation; and c) structured learning hidden units contribution transforms separately conditioned on speaker-identity and speech impairment severity. Experiments conducted on the UASpeech corpus suggest incorporating additional speech impairment severity into state-of-the-art hybrid DNN, E2E Conformer and pre-trained Wav2vec 2.0 ASR systems produced statistically significant word error rate (WER) reductions up to ??\% (??\% relative) on the UASpeech test set of 16 dysarthric speakers. Using the best system the lowest published WER of 17.82\% (??.??\% on Very Low intelligibility) was obtained.}

A key challenge in dysarthric speech recognition is the speaker-level diversity attributed to both speaker-identity associated factors such as gender, and speech impairment severity. Most prior researches on addressing this issue focused on using speaker-identity only. To this end, this paper proposes a novel set of techniques to use both severity and speaker-identity in dysarthric speech recognition: a) multitask training incorporating severity prediction error; b) speaker-severity aware auxiliary feature adaptation; and c) structured LHUC transforms separately conditioned on speaker-identity and severity. Experiments conducted on UASpeech suggest incorporating additional speech impairment severity into state-of-the-art hybrid DNN, E2E Conformer and pre-trained Wav2vec 2.0 ASR systems produced statistically significant WER reductions up to 4.78\% (14.03\% relative). Using the best system the lowest published WER of 17.82\% (51.25\% on very low intelligibility) was obtained on UASpeech.

%Automatic recognition of dysarthric speech is challenging. Speaker-level heterogeneity among dysarthric speakers can be attributed to both factors related to speaker-identity, such as gender or accent, and speech impairment severity. However, the majority of prior researches on addressing such heterogeneity focused on using speaker-identity only. To this end, this paper investigates a novel set of techniques to utilize speech impairment severity for dysarthric speech recognition, including the use of multitask training cost interpolation, auxiliary feature adaptation and structural LHUC transforms. Experiments conducted on the UASpeech corpus suggest incorporating speech impairment severity into state-of-the-art hybrid DNN, E2E Conformer and SSL pre-trained Wav2vec 2.0 ASR systems all produce statistically significant improvement. Combining the best performing systems via two pass rescoring gives an overall word error rate of 17.82\% on the UASpeech test set of 16 dysarthric speakers.}
\end{abstract}


\section{Introduction}
\label{sec:intro}
In spite of the rapid development of automatic speech recognition (ASR) techniques targeting normal speech in recent decades, accurate recognition of dysarthric speech remains a highly challenging task to date~\cite{christensen2012comparative, kim2013dysarthric,sehgal2015model,joy2018improving,yu2018development,hu2019cuhk,xiong2019phonetic,liu2020exploiting,geng2020investigation,harvill2021synthesis,geng2021spectro,xie2021variational,liu2021recent,jin2021adversarial,deng2021bayesian,hu2021neural,geng2022fly,jin2022personalized,mulfari2022exploring,hu2022exploiting,hu2022exploit,tobin2022personalized,baskar2022speaker,hernandez2022cross,geng2022spectro,yue2022acoustic,jin2022adversarial,hu2022exploring}. 
%% The underlying causes of dysarthria include a wide range of neural conditions related to speech control, such as amyotrophic lateral sclerosis, cerebral palsy, traumatic brain injuries, and stroke~\cite{lanier2010speech}. Even though the speech quality of such speakers is degraded, their difficulties in using a keyboard, mouse or touch screen based user interfaces make voice-based assistive technologies more natural alternatives. ASR technologies tailored to their needs can improve their quality of life and social inclusion.
Dysarthric speech imposes several challenges to current deep learning based ASR techniques prevailingly targeting normal speech. These include: a) data scarcity due to the difficulty in collecting a large amount of such speech from dysarthric speakers who often suffer from physical disabilities and mobility issues; b) mismatch against healthy speech; and c) large diversity among dysarthric speakers, when sources of variability commonly found in normal speech including accent or gender are aggregated with those over speech impairment severity. For example, dysarthric speakers of very low speech intelligibility exhibit clearer patterns of articulatory imprecision, decreased volume and clarity, increased dysfluencies, slower speaking rate and changes in pitch~\cite{kent2000dysarthrias}, while those diagonalized with mid or high speech intelligibility are closer to normal speakers. Such heterogeneity further increases the mismatch against normal speech and the difficulty in both speaker-independent (SI) ASR system development using limited impaired speech data and fine-grained personalization to individual users’ data~\cite{sehgal2015model,geng2022spectro,kodrasi2020spectro}

So far the majority of prior researches to address the dysarthric speaker level diversity have been focused on using speaker-identity only either in speaker-dependent (SD) data augmentation~\cite{xiong2019phonetic,geng2020investigation,liu2021recent,jin2021adversarial,jin2022personalized,jin2022adversarial}, or in speaker adapted or dependent ASR system development~\cite{christensen2012comparative,sehgal2015model,bhat2016recognition,kim2017regularized,joy2018improving,xiong2019phonetic,takashima2020two,liu2021recent,geng2021spectro,xie2021variational,tobin2022personalized,mulfari2022exploring,geng2022spectro}. In contrast, very limited prior researches have used speech impairment severity information for dysarthric speech recognition. Dysarthria severity-dependent GMM-HMM state distributions were proposed in~\cite{kim2013dysarthric}. Severity-dependent MLLR and CMLLR adaptations were studied in the context of GMM-HMM models in~\cite{mustafa2014severity}. Severity-dependent tempo perturbation of dysarthric speech to match healthy speech was investigated in~\cite{bhat2016improving}. Speech impairment severity was also used in multi-speaker text-to-speech (TTS) systems to control pitch, energy and duration when synthesizing additional dysarthric training speech data for ASR system development~\cite{soleymanpour2022synthesizing}. However, there is a notable lack of solutions that can account for the dysarthric speech data heterogeneity that is attributed to both speaker-identity and speech impairment severity. In particular, the use of speech impairment severity in current end-to-end (E2E) and pre-trained ASR systems has been rarely visited.

% From professor:
% 1) There is no clear boundary between speaker adaptated and dependent systems, in the sense that, model based adaptation will use SD parameters, while speaker dependent systems have all parameters tuned for the targeted speaker.
% 2) For auxiliary feature based adaptation, the situation is different since no specially designed network parameters are included.

To this end, this paper investigates a novel set of techniques to incorporate speech impairment severity into state-of-the-art hybrid DNN~\cite{liu2021recent}, end-to-end Conformer~\cite{gulati2020conformer} and self-supervised learning (SSL) based pre-trained Wav2vec 2.0~\cite{baevski2020wav2vec} ASR systems. These include the use of: a) multi-task~\cite{caruana1997multitask} training cost interpolation between the ASR loss and speech impairment severity prediction error; b) spectral basis embedding (SBE)\cite{geng2021spectro, geng2022spectro} based speaker-severity aware adaptation features that are trained to simultaneously predict both speaker-identity and impairment severity; and c) structured learning hidden units contribution (LHUC)~\cite{swietojanski2016learning} transforms that are separately conditioned on speaker-identity and impairment severity. These are used to facilitate both speaker-severity adaptive training of ASR systems and their test-time unsupervised adaptation to both factors of variability. Learning both speech impairment severity and speaker-identity serves as a dual-purpose solution. First, it allows speaker and impairment severity invariant ``canonical'' ASR systems to be constructed. Second, these two sources of variability can be flexibly factored in and combined for fine-grained adaptation to diverse dysarthric speakers.

Experiments were conducted on the largest available and most widely used UASpeech~\cite{kim2008dysarthric} dysarthric speech dataset. Experimental results suggest the incorporation of speech impairment severity produced statistically significant~\cite{pallet1990tools} word error rate (WER) reductions up to 3.95\%, 4.78\% and 4.37\% absolute (12.56\%, 14.03\% and 16.53\% relative) for hybrid DNN, E2E Conformer and cross-domain fine-tuned Wav2vec 2.0 models. Modeling both severity and speaker-identity produced further improvements. The lowest published WER of 17.82\% (51.25\% and 17.41\% on very low and low intelligibility) was obtained on the UASpeech test set of $16$ dysarthric speakers by combining the best-performing hybrid DNN, E2E Conformer and fine-tuned Wav2vec 2.0 systems via two pass rescoring~\cite{cui2022two}. 

The main contributions of the paper are summarized below: 

1) To the best of our knowledge, this paper presents the first work of systematically incorporating speech impairment severity into hybrid DNN, E2E Conformer and cross-domain fine-tuning of Wav2vec 2.0 pre-trained ASR models for dysarthric speech recognition. A set of novel techniques and recipe configurations were proposed to learn both speech impairment severity and speaker-identity when constructing and personalizing these systems. In contrast, prior researches mainly focused on using speaker-identity only in speaker-dependent data augmentation~\cite{xiong2019phonetic,geng2020investigation,liu2021recent,jin2021adversarial,jin2022personalized,jin2022adversarial} and speaker adapted or dependent ASR system development~\cite{christensen2012comparative,sehgal2015model,bhat2016recognition,kim2017regularized,joy2018improving,xiong2019phonetic,takashima2020two,liu2021recent,geng2021spectro,baskar2022speaker,tobin2022personalized,mulfari2022exploring,geng2022spectro}. Very limited prior researches utilized speech impairment severity information~\cite{kim2013dysarthric, mustafa2014severity, bhat2016improving,soleymanpour2022synthesizing,geng2021spectro,geng2022spectro}. None of these was conducted in the context of fine-tuning state-of-the-art pre-trained ASR models for dysarthric speech recognition, as considered in this paper. 

% \textcolor{red}{Compared with constructing SD systems for specific known speakers, developing SI model with our proposed personalization techniques can be better generalized to unseen users and meet the practical needs.}
% the above is not proper, since for UASpeech we don't have test speakers that is not in the traininhg set

2) The final system combining the best-performing hybrid DNN, E2E Conformer and fine-tuned Wav2vec 2.0 systems via two pass rescoring gave an overall WER of 17.82\% on the UASpeech test set of $16$ dysarthric speakers. This is the best performance reported so far on UASpeech as far as we know.

% \textbf{Paragraph 1}: Dysarthric speech is challenging. Causes of Dysarthria. Meaning of speech recognition based assisitive techonology.

% \noindent \textbf{Paragraph 2}: Significant mismatch for speakers with  pathological severity.

% \noindent \textbf{Paragraph 3}: Limit previous researches on building recognition systems that incorporates pathological severity information.

% \noindent \textbf{Paragraph 4}: Incorporating pathological severity in hybrid DNN and E2E conformer \& Wav2vec2 pre-trained systems are systematically investigated in this paper.

% \noindent \textbf{Paragraph 5}: Summary of contribution of this paper.

\vspace{-0.5em}
\section{Incorporation of Speech Impairment Severity into Hybrid ASR Systems}
\label{sec:hybrid}
\begin{figure}[ht]
  \centering
  \vspace{-0.2cm} 
  \setlength{\abovecaptionskip}{0.2cm}   %调整图片标题与图距离
  \setlength{\belowcaptionskip}{-0.2cm}   %调整图片标题与下文距离
  \includegraphics[scale=0.64]{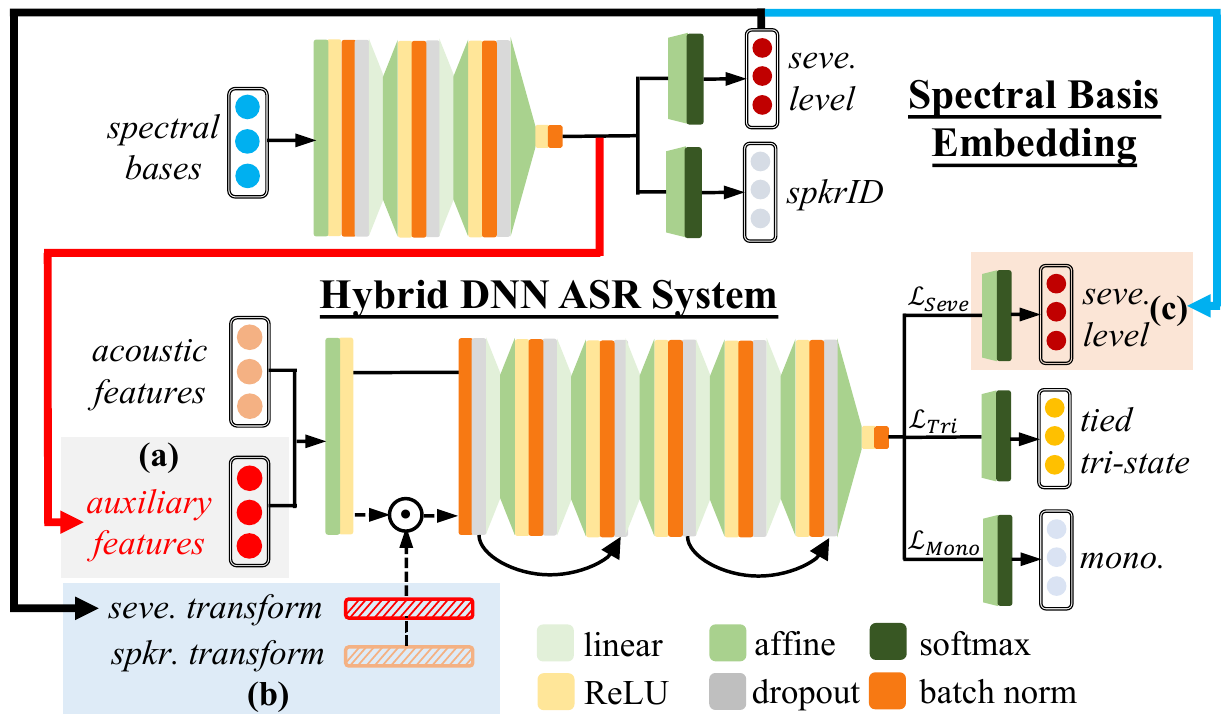}
  \caption{Incorporating speech impairment severity into the hybrid DNN system at \textbf{(a)} model input via speaker-severity aware auxiliary features, \textbf{(b)} structured speaker-severity LHUC transforms, and \textbf{(c)} model output via multitask training with severity prediction error. The upper network using spectral bases as inputs serves as a dual-purpose solution of generating speaker-severity aware auxiliary features and automatically assessing the severity levels of test speakers. ``seve.'', ``spkrID'' and ``mono.'' are short for severity, speaker ID and monophone.}
  \label{fig:hybrid}
\end{figure}

In this section, we propose a novel set of techniques to incorporate speech impairment severity into the hybrid DNN~\cite{liu2021recent} systems (Fig.~\ref{fig:hybrid}). These include the use of: 1) speaker-severity aware auxiliary features serving as front-ends; 2) structured learning hidden unit contributions (LHUC) transforms separately conditioned on speaker-identity and severity; and 3) multitask learning incorporating severity prediction error.

\subsection{Speaker-Severity Aware Auxiliary Features}
The underlying variability of dysarthric speech manifesting in changes of spectral envelope, volume reduction, imprecise articulation and breathy or horse voices, can be modeled via disentangling the speech spectrum into time-invariant and time-variant subspaces~\cite{kodrasi2020spectro,geng2021spectro} learned in a supervised manner~\cite{geng2021spectro,geng2022spectro}. The resulting spectral basis deep embedding (SBE) features are more effective in encoding latent attributes of impaired speech~\cite{geng2022spectro} than classical speaker embeddings such as iVectors~\cite{senior2014improving} and xVectors~\cite{snyder2018x}. Hence, we adopt SBE features as speaker-severity aware auxiliary inputs to the hybrid DNN systems to incorporate both speaker-identity and speech impairment severity into the acoustic front-ends.

% Following~\cite{geng2021spectro}, singular value decomposition (SVD) is conducted on the speech spectrum to generate decomposed spectral and temporal subspaces. Top-$2$ bases from the spectral subspace, which is more related to the underlying speech impairment severity, are fed into a $3$-layer DNN embedding network (Fig.~\ref{fig:hybrid} upper) with severity levels and speaker IDs as targets~\cite{geng2022spectro}. During training, spectral bases of the training speakers are fed into the embedding network, and the $25$-dim compact features extracted from the last bottleneck layer are used as the auxiliary features (Fig.~\ref{fig:hybrid} red bold line). During test time evaluation, spectral bases of the test speaker are fed into the trained embedding network to generate auxiliary features and automatically assess the speaker's speech impairment severity level.

Following~\cite{geng2021spectro}, singular value decomposition (SVD) is first conducted on the speech spectrum to generate decomposed spectral and temporal subspaces. Top-$2$ bases from the spectral subspace, which are more related to the underlying speech impairment severity, are then fed into a $3$-layer DNN embedding network (Fig.~\ref{fig:hybrid} upper) with severity levels and speaker IDs as targets~\cite{geng2022spectro}. During training, the $25$-dim compact features extracted from the last bottleneck layer serve as the auxiliary features (Fig.~\ref{fig:hybrid} red bold line). During test time evaluation, spectral bases of impaired speech utterances are fed into the trained embedding network to generate auxiliary features, and also predict the speech impairment severity level for each test speaker.

\subsection{Structured Speaker-Severity LHUC Adaptation}
LHUC~\cite{swietojanski2016learning} adaptation can model the large variability among dysarthric speakers~\cite{geng2020investigation, liu2021recent}, with speaker-dependent LHUC transforms applied to DNN hidden layers. Inspired by the factorized LHUC representations respectively modeling the speaker and the acoustic environment~\cite{swietojanski2016learning}, we propose structured speaker-severity LHUC adaptation separately conditioned on speaker-identity ($\bm{A}_{spkr}^{s}$) and speech impairment severity ($\bm{A}_{seve}^{s}$). Such LHUC transforms are further restricted as diagonal matrices to reduce the number of parameters~\cite{swietojanski2016learning}, equivalent to using scaling vectors to modify the amplitudes of the ReLU activation in the first layer of the DNN (Fig.~\ref{fig:hybrid} (b)).

Let $\bm{r}_{spkr}^{s}$ and  $\bm{r}_{seve}^{s}$ denote the speaker-dependent and severity-dependent scaling vectors for speaker $s$. The hidden layer output adapted to speaker $s$ is given as:

\vspace{-1em}
\begin{equation}
\label{eq:LHUC}
\begin{split}
    \setlength{\abovedisplayskip}{-3pt}
    \setlength{\belowdisplayskip}{-3pt}
    \bm{h}^{s} & = diag(\bm{A}_{spkr}^{s})diag(\bm{A}_{seve}^{s})\bm{h} \\
               & = \xi(\bm{r}_{spkr}^{s}) \odot \xi(\bm{r}_{seve}^{s}) \odot \bm{h}
\end{split}
\end{equation}
\vspace{-1em}

where $\odot$ is the Hadamard product and $\xi(\cdot)$ is the element-wise $2 \times sigmoid(\cdot)$. During unsupervised test time adaptation, severity levels of test speakers are automatically assessed using the spectral basis embedding network (Fig.~\ref{fig:hybrid}  black bold line).

% factor in (combine) vs. factor out
\subsection{Multitask Learning}
Multitask learning (MTL)~\cite{caruana1997multitask} is proven to be helpful in improving the generalization ability of each task~\cite{pironkov2016multi}. To this end, incorporating severity prediction error in the training criteria of the DNN ASR systems can help produce a neutral, canonical model that is invariant to speech impairment severity. An interpolation between the cross entropy (CE) loss on the frame-level tied triphone states (tri-states) ($\mathcal{L}_{Tri}$),  monophone alignments ($\mathcal{L}_{Mono}$) and speech impairment severity levels ($\mathcal{L}_{Seve}$) is utilized as the multitask loss function, given as:

\vspace{-1.5em}
\begin{equation}
\label{eq:MTL_DNN}
    \mathcal{L}_{MTL_{DNN}} = {\omega_1}{\cdot}\mathcal{L}_{Tri}+{\omega_2}{\cdot}\mathcal{L}_{Mono}+{\omega_3}{\cdot}\mathcal{L}_{Seve}\footnote{Weights are empirically set as $\omega_1=\omega_2=\omega_3=\frac{1}{3}$.}
\end{equation}
\vspace{-1.5em}

Incorporating $\mathcal{L}_{Mono}$ in the loss reduces the risk of over-fitting to unreliable frame-level tri-states,
%% attributed to speech impairment, 
while incorporating $\mathcal{L}_{Seve}$ increases the generalization ability of the model to test speakers with diverse severity levels. During test time adaptation, the unsupervised severity label of the test speaker is derived from automatic assessment via the spectral basis embedding network (Fig.~\ref{fig:hybrid} blue bold line).

\subsection{Severity-Dependent Systems with KLD Regularization}
Motivated by~\cite{kim2013dysarthric} where separate GMM-HMM acoustic models are trained for each severity group, we develop severity-dependent DNN systems as an ablation study. To prevent overfitting to the limited amount of severity-dependent data, we further introduce Kullback-Leibler divergence (KLD) based regularized adaptation~\cite{yu2013kl} by adding the KLD between the output distributions of the unadapted SI and the severity-dependent models into the training cost. This is given as:

\vspace{-1em}
\begin{equation}
\label{eq:KLD}
    \mathcal{L}_{KLD} = (1-\lambda)\mathcal{L} + \lambda \frac{1}{N}\sum_{t=1}^{N}\sum_{y=1}^{S} p^{SI}(y|x_{t})\log p(y|x_{t})
\end{equation}
\vspace{-1em}

% where $\mathcal{L}$ is the standard CE loss to train a hybrid DNN model, $N$ is the number of training samples and $S$ is the number of tri-states. $p^{SI}(\cdot)$ is the posterior probability of the SI model. $\lambda$ is a tunable regularization weight.

where $\mathcal{L}$ is the standard CE loss to train a DNN model, $N$ is the number of training samples and $S$ is the number of tri-states. $p^{SI}(\cdot)$ is the output distribution of the SI model. $\lambda$ is a tunable regularization weight empirically set to $0.5$ in the experiments.

\vspace{-0.5em}
\section{Incorporation of Speech Impairment Severity into E2E Systems}
\label{sec:E2E}
% In this section, we discuss the incorporation of speech impairment severity into the training of start-of-the-art E2E Conformer systems and fine-tuning of the self-supervised learning (SSL) based Wav2vec 2.0 pre-trained systems.

\begin{figure*}[ht]
  \centering
  \vspace{-0.2cm} 
  \setlength{\abovecaptionskip}{0.2cm}   %调整图片标题与图距离
  \setlength{\belowcaptionskip}{-0.5cm}   %调整图片标题与下文距离
  \includegraphics[scale=0.51]{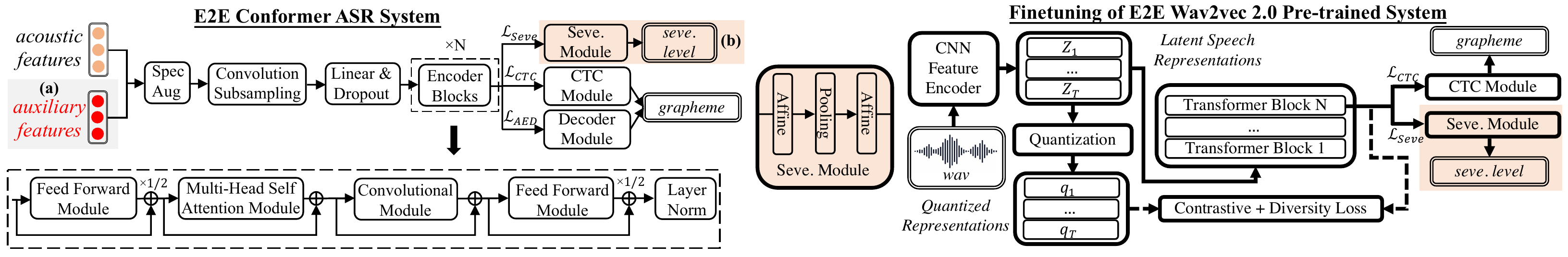}
  \caption{Incorporating speech impairment severity into the E2E Conformer system (left) using \textbf{(a)} speaker-severity aware auxiliary features and \textbf{(b)} multitask training with severity prediction error. Speech impairment severity is incorporated during the fine-tuning of the E2E Wav2vec 2.0 pre-trained system (right) via multitask learning. Components of the severity module are shown in the middle.}
  \label{fig:E2E}
\end{figure*}

\subsection{E2E Conformer Systems}
We incorporate speech impairment severity into E2E Conformer systems through two approaches: 1) using speaker-severity aware auxiliary features (Fig.~\ref{fig:E2E} left (a)) and 2) adding a CE-based severity prediction error $L_{Seve}$ into the training criteria (Fig.~\ref{fig:E2E} left (b)). The MTL training cost is given as:

\vspace{-1.5em}
\begin{equation}
\label{eq:MTL_CONF}
    \mathcal{L}_{MTL_{CONF}} = {\alpha_1}{\cdot}\mathcal{L}_{CTC}+{\alpha_2}{\cdot}\mathcal{L}_{AED}+{\alpha_3}{\cdot}\mathcal{L}_{Seve}\footnote{Weights are empirically set as $\alpha_1=\alpha_2=\alpha_3=\frac{1}{3}$.}
\end{equation}
\vspace{-1.5em}

where $L_{CTC}$ and $L_{AED}$ denote the Connectionist Temporal Classifcation loss~\cite{graves2006connectionist} and the attention-based encoder-decoder loss~\cite{vaswani2017attention}, respectively.

\subsection{Fine-tuning of SSL Wav2vec 2.0 pre-trained Systems}
Speech impairment severity is incorporated by adding CE-based severity prediction error in the training cost during the cross-domain fine-tuning of SSL-based Wav2vec 2.0 pre-trained systems (Fig.~\ref{fig:E2E} right). The MTL criterion is given as:

\vspace{-1em}
\begin{equation}
\label{eq:MTL_W2V}
    \mathcal{L}_{MTL_{W2V}} = {\beta_1}{\cdot}\mathcal{L}_{CTC}+{\beta_2}{\cdot}\mathcal{L}_{Seve}\footnote{Weights are empirically set as $\beta_1=\beta_2=\frac{1}{2}$.}
\end{equation}
\vspace{-1em}

% The aforementioned approaches of incorporating 

\vspace{-0.5em}
\section{Experiments and Results}
\label{sec:exp}
\subsection{Task Description}
The UASpeech~\cite{kim2008dysarthric} corpus is the largest publicly available and widely used dysarthric speech dataset. It is an isolated word recognition task containing $103$h speech from $16$ dysarthric and $13$ control speakers. The dysarthric speakers are further divided into speech intelligibility subgroups ``very low'', ``low'', ``mid'' and ``high''. For each speaker, the data is split into three blocks B1, B2 and B3, each with the same $155$ common words and a different set of $100$ uncommon words. The training set includes the data from B1 and B3 of all $29$ speakers ($69.1$h), while the test set includes the data from B2 of all $16$ dysarthric speakers ($22.6$h, excluding speech from control speakers). Silence stripping using an HTK~\cite{young2002htk} trained GMM-HMM system~\cite{liu2021recent} produces a $30.6$h training set ($99195$ utt.) and a $9$h test set ($26520$ utt.). Data augmentation via speed perturbation~\cite{geng2020investigation} produces a $130.1$h augmented training set ($399110$ utt.). As E2E systems are sensitive to the training data coverage (Sys.2 vs. Sys.1 in Table~\ref{tab:recog-UASpeech-Conformer}-\ref{tab:recog-UASpeech-W2V}), B2 data of the $13$ control speakers are further folded in during the training of Conformer and fine-tuning of Wav2vec 2.0 systems following~\cite{hernandez2022cross}. This produces a $40$h unaugmented ($122392$ utt.) and a $190$h augmented training set ($538292$ utt.). 

\subsection{Experiment Setup}
% \noindent{\textbf{Hybrid DNN systems:}} The $7$-layer hybrid DNN systems were implemented using Kaldi~\cite{povey2011kaldi} following ~\cite{liu2021recent}. The inputs were $80$-dim filter-bank (FBK) + $\Delta$ features, optionally plus $25$-dim speaker-severity aware auxiliary features. A uniform language model (LM) was used in decoding~\cite{christensen2012comparative}. 

\noindent{\textbf{Hybrid DNN systems:}} The $7$-layer hybrid DNN systems were implemented using Kaldi~\cite{povey2011kaldi} following ~\cite{liu2021recent}. The inputs were $80$-dim filter-bank (FBK) + $\Delta$ features, optionally plus $25$-dim speaker-severity aware auxiliary features.

\noindent{\textbf{E2E Conformer systems:}} The graphemic Conformer systems were implemented via ESPnet~\cite{watanabe2018espnet}\footnote{$12$ encoder layers + $12$ decoder layers, feed-forward dim = $2048$, $4$ attention heads of $256$ dimensions.}. The inputs were $80$-dim FBK + $\Delta$ features, optionally plus $25$-dim auxiliary features. 

\noindent{\textbf{Wav2vec 2.0 pre-trained systems:}} The Wav2vec 2.0 pre-trained systems (Fig.~\ref{fig:E2E} right) contained three components: 1) a speech feature encoder with CNN convolution blocks, 2) a contextual transformer network and 3) a quantization module. The inputs were raw speech waves. During pre-training, an interpolation between the contrastive and the diversity loss was used to train the model. During fine-tuning, the CTC loss, optionally with an interpolation of the severity prediction error, was used. The large Wav2Vec 2.0 model\footnote{https://huggingface.co/facebook/wav2vec2-large-960h-lv60} pre-trained using $60k$ hours of Libri-light data and fine-tuned using $960$h Librispeech data was used as the base for cross-domain fine-tuning on UASpeech.

\subsection{Result Analysis}
Table~\ref{tab:recog-UASpeech-hybrid-1} shows the performance comparison\footnote{A matched pairs sentence-segment word error (MAPSSWE) based statistical signiﬁcance test~\cite{pallet1990tools} was done at signiﬁcance level $\alpha=0.05$.} on the $30.6$h training set between separately trained severity-dependent DNN models, speaker-independent (SI) models, and models with severity incorporated via auxiliary features, multitask learning or structured LHUC transform. Several trends can be observed: \textbf{1)} The SI system (Sys.3) outperforms the severity-dependent systems (Sys.1-2) and thus is chosen as the baseline for further experiments. \textbf{2)} Incorporating severity via auxiliary features, multitask learning, or structured LHUC transforms into the systems (Sys.4-6) all produce statistically significant improvements over the SI baseline (Sys.3). Combining two approaches (Sys.7-9) produce WER reductions up to {\bf 3.95\% absolute (12.56\% relative)} over the baseline (Sys.7 vs. 3). Combining all three approaches (Sys.10) leads to no further improvement. \textbf{3)} When further combined with LHUC-SAT speaker adaptation, similar trends are observed (Sys.12-14,15-17,18 vs. Sys.11). A similar set of experiments on the $130$h augmented training set serve to further evaluate the above techniques proposed for modeling severity, as shown in Table~\ref{tab:recog-UASpeech-hybrid-2}.
%% The strategies producing competitive performance are further validated using the augmented training sets (Table~\ref{tab:recog-UASpeech-hybrid-2}).

\vspace{-0.5em}
\begin{table}[ht]
    \caption{Performance of incorporating severity into DNN models on the 16 UASpeech dysarthric speakers using the 30.6h training set.  ``Aux. Feat.'' and ``Trn. Tar.'' denote auxiliary feature and training target. ``Seve.'' and ``Intel.'' stand for severity and intelligibility. $^\dag$ denotes a statistically significant improvement ($\alpha=0.05$) is obtained over the comparable baseline systems without using severity information (Sys.3,11).}
    \label{tab:recog-UASpeech-hybrid-1}
    \Large
    \centering
    \vspace{-0.5em}
    \setlength{\abovecaptionskip}{0.05cm}
    \renewcommand\arraystretch{1.0}
    \renewcommand\tabcolsep{2.0pt}
    \scalebox{0.39}{\begin{tabular}{c|c|c|c|ccc|c|cccc|cc|c}
    \hline\hline  
        \multirow{3}{*}{Sys.} & 
        \multirow{3}{*}{Model} & 
        \multirow{3}{*}{KLD} & 
        \multirow{3}{*}{\#Hrs} &
        \multicolumn{3}{c|}{Impair. Seve.} &
        \multirow{3}{*}{\tabincell{c}{LHUC\\SAT\\(Spkr)}} &
        \multicolumn{7}{c}{WER\%} \\
    \cline{5-7}\cline{9-15}
        & & & &
        \multirow{2}{*}{\tabincell{c|}{Aux.\\Feat.}} & 
        \multirow{2}{*}{\tabincell{c|}{Trn.\\Tar.}} & 
        \multirow{2}{*}{\tabincell{c}{LHUC\\(Seve.)}} &
        &
        \multicolumn{4}{c|}{Intel. Subgroup} &
        \multirow{2}{*}{Unseen} & 
        \multirow{2}{*}{Seen} &
        \multirow{2}{*}{All} \\ 
    \cline{9-12}
        & & & & & & & & VL & L & M & H & & & \\
    \hline\hline
        1 & \multirow{2}{*}{\tabincell{c}{Seve.\\DNN}} & \xmark & \multirow{3}{*}{30.6} & \multirow{2}{*}{\xmark} & \multirow{2}{*}{\xmark} & 
        \multirow{2}{*}{\xmark} & \multirow{2}{*}{\xmark} & 70.96 & 35.29 & 26.35 & 11.05 & 47.47 & 23.60 & 32.96 \\
        2 & & \cmark & & & & & & 70.14 & 34.84 & 25.76 & 10.52 & 46.67 & 23.15 & 32.37 \\
    \cline{1-3}\cline{5-15}
        3 & DNN & / & & \xmark & \xmark & \xmark & \xmark & 69.82 & 32.61 & 24.53 & 10.40 & 46.18 & 21.94 & 31.45 \\
    \cline{1-15}
        4 & \multirow{7}{*}{DNN} & \multirow{7}{*}{/} & \multirow{7}{*}{30.6} & \cmark & \xmark & \xmark & \xmark & 64.43$^\dag$ & 29.71$^\dag$ & 19.84$^\dag$ & 8.57$^\dag$ & 42.82$^\dag$ & 18.51$^\dag$ & \textbf{28.05$^\dag$} \\
        5 & & & & \xmark & \cmark & \xmark & \xmark & 69.09$^\dag$ & 29.35$^\dag$ & 21.01$^\dag$ & 9.78$^\dag$ & 43.43$^\dag$ & 20.62$^\dag$ & 29.57$^\dag$ \\
        6 & & & & \xmark & \xmark & \cmark & \xmark & 67.21$^\dag$ & 29.67$^\dag$ & 20.82$^\dag$ & 9.07$^\dag$ & 43.16$^\dag$ & 19.83$^\dag$ & 28.98$^\dag$ \\
    \cline{1-1}\cline{5-15}
        7 & & & & \cmark & \cmark & \xmark & \xmark & \textbf{63.83$^\dag$} & \textbf{28.00$^\dag$} & \textbf{19.80$^\dag$} & \textbf{8.68$^\dag$} & \textbf{41.71$^\dag$} & \textbf{18.33$^\dag$} & \textbf{27.50$^\dag$} \\
        8 & & & & \xmark & \cmark & \cmark & \xmark & 67.18$^\dag$ & 30.18$^\dag$ & 21.27$^\dag$ & 9.00$^\dag$ & 43.57$^\dag$ & 19.87$^\dag$ & 29.17$^\dag$ \\
        9 & & & & \cmark & \xmark & \cmark & \xmark & 64.56$^\dag$ & 28.97$^\dag$ & 19.01$^\dag$ & 8.53$^\dag$ & 42.49$^\dag$ & 18.17$^\dag$ & 27.71$^\dag$ \\
    \cline{1-1}\cline{5-15}
        10 & & & & \cmark & \cmark & \cmark & \xmark & 64.52$^\dag$ & 28.77$^\dag$ & 19.90$^\dag$ & 8.31$^\dag$ & 42.33$^\dag$ & 18.33$^\dag$ & 27.75$^\dag$ \\
    \hline\hline 
        11 & \multirow{8}{*}{DNN} & \multirow{8}{*}{/} & \multirow{8}{*}{30.6} & \xmark & \xmark & \xmark & \cmark & 64.39 & 29.88 & 20.27 & 8.95 & 42.32 & 19.23 & 28.29 \\
    \cline{1-1}\cline{5-15}
        12 & & & & \cmark & \xmark & \xmark & \cmark & 63.40$^\dag$ & 28.90$^\dag$ & 18.64$^\dag$ & 8.13$^\dag$ & 41.79$^\dag$ & 17.84$^\dag$ & \textbf{27.24$^\dag$} \\
        13 & & & & \xmark & \cmark & \xmark & \cmark & 65.70 & 28.40$^\dag$ & 19.43$^\dag$ & 9.06 & 42.24 & 18.91$^\dag$ & 28.06  \\
        14 & & & & \xmark & \xmark & \cmark & \cmark & 65.09 & 28.83$^\dag$ & 19.58$^\dag$ & 8.56$^\dag$ & 42.85 & 18.25$^\dag$ & 27.90$^\dag$  \\
    \cline{1-1}\cline{5-15}
        15 & & & & \cmark & \cmark & \xmark & \cmark & 64.52 & 28.10$^\dag$ & 18.56$^\dag$ & 8.01$^\dag$ & 41.35$^\dag$ & 18.08$^\dag$ & 27.21$^\dag$ \\
        16 & & & & \xmark & \cmark & \cmark & \cmark & 64.49 & 29.29$^\dag$ & 18.15$^\dag$ & 8.52$^\dag$ & 42.06 & 18.28$^\dag$ & 27.61$^\dag$ \\
        17 & & & & \cmark & \xmark & \cmark & \cmark & \textbf{63.65$^\dag$} & \textbf{28.11$^\dag$} & \textbf{18.31$^\dag$} & \textbf{8.41$^\dag$} & \textbf{41.99$^\dag$} & \textbf{17.52$^\dag$} & \textbf{27.12$^\dag$} \\
    \cline{1-1}\cline{5-15}
        18 & & & & \cmark & \cmark & \cmark & \cmark & 64.86 & 28.16$^\dag$ & 19.29$^\dag$ & 8.23$^\dag$ & 42.13 & 18.08$^\dag$ & 27.52$^\dag$ \\
    \hline\hline 
    \end{tabular}}
\end{table}

\vspace{-1.5em}

\begin{table}[ht]
    \caption{Performance of incorporating severity into DNN models on UASpeech using the 130.1h or 190h augmented training set. $^\dag$ denotes a statistically significant improvement ($\alpha=0.05$) over the baseline systems (Sys.3,7,11,13).}
    \label{tab:recog-UASpeech-hybrid-2}
    \Large
    \centering
    \vspace{-0.5em}
    \setlength{\abovecaptionskip}{0.05cm}
    \renewcommand\arraystretch{1.0}
    \renewcommand\tabcolsep{2.0pt}
    \scalebox{0.39}{\begin{tabular}{c|c|c|c|ccc|c|cccc|cc|c}
    \hline\hline  
        \multirow{3}{*}{Sys.} & 
        \multirow{3}{*}{Model} & 
        \multirow{3}{*}{KLD} & 
        \multirow{3}{*}{\#Hrs} &
        \multicolumn{3}{c|}{Impair. Seve.} &
        \multirow{3}{*}{\tabincell{c}{LHUC\\SAT\\(Spkr)}} &
        \multicolumn{7}{c}{WER\%} \\
    \cline{5-7}\cline{9-15}
        & & & &
        \multirow{2}{*}{\tabincell{c|}{Aux.\\Feat.}} & 
        \multirow{2}{*}{\tabincell{c|}{Trn.\\Tar.}} & 
        \multirow{2}{*}{\tabincell{c}{LHUC\\(Seve.)}} &
        &
        \multicolumn{4}{c|}{Intel. Subgroup} &
        \multirow{2}{*}{Unseen} & 
        \multirow{2}{*}{Seen} &
        \multirow{2}{*}{All} \\ 
    \cline{9-12}
        & & & & & & & & VL & L & M & H & & & \\
    \hline\hline
        1 & \multirow{2}{*}{\tabincell{c}{Seve.\\DNN}} & \xmark & \multirow{3}{*}{130.1} & \multirow{2}{*}{\xmark} & \multirow{2}{*}{\xmark} & 
        \multirow{2}{*}{\xmark} & \multirow{2}{*}{\xmark} & 66.61 & 29.52 & 22.70 & 9.90 & 43.55 & 20.35 & 29.45 \\
        2 & & \cmark & & & & & & 66.54 & 29.38 & 22.00 & 9.38 & 43.22 & 19.98 & 29.09 \\
    \cline{1-3}\cline{5-15}
        3 & DNN & / & & \xmark & \xmark & \xmark & \xmark & 66.45 & 28.95 & 20.37 & 9.62 & 42.46 & 19.86 & 28.73 \\
    \cline{1-15}
        4 & \multirow{3}{*}{DNN} & \multirow{3}{*}{/} & \multirow{3}{*}{130.1} & \cmark & \xmark & \xmark & \xmark & \textbf{61.55$^\dag$} & \textbf{27.52$^\dag$} & \textbf{17.31$^\dag$} & \textbf{8.22$^\dag$} & \textbf{40.18$^\dag$} & \textbf{17.28$^\dag$} & \textbf{26.26$^\dag$}  \\
        5 & & & & \cmark & \cmark & \xmark & \xmark & 63.83$^\dag$ & 28.00$^\dag$ & 19.80$^\dag$ & 8.68$^\dag$ & 40.63$^\dag$ & 17.54$^\dag$ & 26.60$^\dag$ \\
        6 & & & & \cmark & \xmark & \cmark & \xmark & 62.24$^\dag$ & 27.81$^\dag$ & 17.25$^\dag$ & 8.15$^\dag$ & 40.89$^\dag$ & 17.13$^\dag$ & 26.45$^\dag$ \\
    \hline\hline 
        7 & \multirow{4}{*}{DNN} & \multirow{4}{*}{/} & \multirow{4}{*}{130.1} & \xmark & \xmark & \xmark & \cmark & 62.50 & 27.26 & 18.41 & 8.04 & 40.01 & 17.85 & 26.55   \\
    \cline{1-1}\cline{5-15}
        8 & & & & \cmark & \xmark & \xmark & \cmark & \textbf{59.83$^\dag$} & \textbf{27.16} & \textbf{16.80$^\dag$} & \textbf{7.91} & \textbf{39.55$^\dag$} & \textbf{16.60$^\dag$} & \textbf{25.60$^\dag$} \\
        9 & & & & \cmark & \cmark & \xmark & \cmark & 61.49$^\dag$ & 27.05 & 17.19$^\dag$ & 7.84 & 40.01 & 16.92$^\dag$ & 25.98$^\dag$ \\
        10 & & & & \cmark & \xmark & \cmark & \cmark & 61.21$^\dag$ & 27.62 & 17.60$^\dag$ & 7.80 & 40.47 & 16.88$^\dag$ & 26.13$^\dag$ \\
    \hline\hline
        11 & \multirow{4}{*}{DNN} & \multirow{4}{*}{/} & \multirow{4}{*}{190} & \xmark & \xmark & \xmark & \xmark & 67.41 & 29.17 & 18.90 & 7.48 & 39.50 & 20.56 & 27.99 \\
        12 & & & & \cmark & \xmark & \xmark & \xmark & 61.76$^\dag$ & 27.11$^\dag$ & 15.96$^\dag$ & 6.54$^\dag$ & 36.85$^\dag$ & 17.97$^\dag$ & \textbf{25.37$^\dag$} \\
    \cline{1-1}\cline{5-15} 
        13 & & & & \xmark & \xmark & \xmark & \cmark & 62.60$^\dag$ & 26.65$^\dag$ & 15.80$^\dag$ & 5.64$^\dag$ & 36.98 & 17.43 & 25.10$^\dag$ \\
        14 & & & & \cmark & \xmark & \xmark & \cmark & 62.31 & 25.15$^\dag$ & 13.00$^\dag$ & 4.66$^\dag$ & 33.43$^\dag$ & 17.55 & \textbf{23.78$^\dag$} \\
    \hline\hline 
    \end{tabular}}
\end{table}

\vspace{-1em}

From Table~\ref{tab:recog-UASpeech-hybrid-2}, the SI system (Sys.3) outperforms the separately trained severity-dependent systems (Sys.1-2) on the $130.1$h augmented training set. The best-performing $130.1$h systems with or without standard LHUC SAT are obtained by incorporating severity through auxiliary features (Sys.4,8). Their improvements on the $190$h augmented training set are also consistently observed (Sys. 12 vs. 11, Sys. 14 vs. 13).

\begin{table}[ht]
    \caption{Performance of incorporating severity into E2E Conformer (CONF.) models on UASpeech. $^\dag$ denotes a significant improvement ($\alpha=0.05$) over the baseline (Sys.2).}
    \label{tab:recog-UASpeech-Conformer}
    \Large
    \centering
    \vspace{-0.5em}
    \setlength{\abovecaptionskip}{0.05cm}
    \renewcommand\arraystretch{1.0}
    \renewcommand\tabcolsep{2.0pt}
    \scalebox{0.44}{\begin{tabular}{c|c|c|cc|cccc|cc|c}
    \hline\hline  
        \multirow{3}{*}{Sys.} & 
        \multirow{3}{*}{Model} & 
        \multirow{3}{*}{\#Hrs} &
        \multicolumn{2}{c|}{Impair. Seve.} &
        \multicolumn{7}{c}{WER\%} \\
    \cline{4-12}
        & & &
        \multirow{2}{*}{\tabincell{c|}{Aux.\\Feat.}} & 
        \multirow{2}{*}{\tabincell{c}{Trn.\\Tar.}} & 
        \multicolumn{4}{c|}{Intel. Subgroup} &
        \multirow{2}{*}{Unseen} & 
        \multirow{2}{*}{Seen} &
        \multirow{2}{*}{All} \\ 
    \cline{6-9}
        & & & & & VL & L & M & H & & & \\
    \hline\hline
        1 & \multirow{5}{*}{\tabincell{c}{CONF.}} & 130.1 & \xmark & \xmark & 73.88 & 53.12 & 49.92 & 42.03 & 99.14  & 23.51 & 53.17 \\
    \cline{1-1}\cline{3-12}
        2 & & \multirow{4}{*}{190} & \xmark & \xmark & 65.70 & 40.63 & 33.39 & 9.53 & 57.40 & 19.03 & 34.07 \\
    \cline{1-1}\cline{4-12}
        3 & & & \cmark & \xmark & \textbf{65.18$^\dag$} & \textbf{34.90$^\dag$} & \textbf{24.21$^\dag$} & \textbf{5.00$^\dag$} & \textbf{47.77$^\dag$} & \textbf{17.20$^\dag$}  & \textbf{29.19$^\dag$}  \\
        4 & & & \xmark & \cmark & 66.95 & 39.08$^\dag$  & 29.78$^\dag$  & 8.03$^\dag$  & 56.01$^\dag$ & 17.72$^\dag$ & 32.74$^\dag$  \\
        5 & & & \cmark & \cmark & 68.03 & 36.36$^\dag$  & 23.29$^\dag$  & 4.76$^\dag$  & 47.16$^\dag$  & 18.79 & 29.91$^\dag$  \\   
    \hline\hline 
    \end{tabular}}
\end{table}

\begin{table}[ht]
    \caption{Performance of finetuing Wav2vec 2.0 (W2V.) pre-trained model on UASpeech. $^\dag$ denotes a statistically significant improvement ($\alpha=0.05$) over the baseline (Sys.2).}
    \label{tab:recog-UASpeech-W2V}
    \Large
    \centering
    \vspace{-0.5em}
    \setlength{\abovecaptionskip}{0.05cm}
    \renewcommand\arraystretch{1.0}
    \renewcommand\tabcolsep{2.0pt}
    \scalebox{0.49}{\begin{tabular}{c|c|c|c|cccc|cc|c}
    \hline\hline  
        \multirow{3}{*}{Sys.} & 
        \multirow{3}{*}{Model} & 
        \multirow{3}{*}{\#Hrs} &
        \multirow{3}{*}{\tabincell{c}{Seve.\\Task}} &
        \multicolumn{7}{c}{WER\%} \\
    \cline{5-11}
        & & & &
        \multicolumn{4}{c|}{Intel. Subgroup} &
        \multirow{2}{*}{Unseen} & 
        \multirow{2}{*}{Seen} &
        \multirow{2}{*}{All} \\ 
    \cline{5-8}
        & & & & VL & L & M & H & & & \\
    \hline\hline
        1 & \multirow{3}{*}{\tabincell{c}{W2V.}} & 30.6 & \xmark & 68.25 & 37.81  & 24.00 & 8.13 & 52.25 & 18.29 & 31.61 \\
    \cline{1-1}\cline{3-11}
        2 & & \multirow{2}{*}{40} & \xmark  & 69.04 & 29.86  & 14.75 & 3.68 & 36.52 & 19.92 & 26.44 \\
    \cline{1-1}\cline{4-11}
        3 & & & \cmark & \textbf{59.38$^\dag$} & \textbf{23.91$^\dag$} & \textbf{12.10$^\dag$} & \textbf{2.91$^\dag$} & \textbf{33.27$^\dag$} & \textbf{14.85$^\dag$} & \textbf{22.07$^\dag$} \\
    \hline\hline 
    \end{tabular}}
    \vspace{-1em}
\end{table}

Tables~\ref{tab:recog-UASpeech-Conformer} and \ref{tab:recog-UASpeech-W2V} show the performance of incorporating severity into Conformer and Wav2vec 2.0 systems\footnote{Using the $30.6$h and the $130.1$h training set on UASpeech produce comparable performance when fine-tuning Wav2vec 2.0 models~\cite{hu2022exploring}.}. Using severity in Conformer produces statistically significant WER reductions up to {\bf 4.78\% absolute (14.03\% relative)} (Table~\ref{tab:recog-UASpeech-Conformer}, Sys. 3 vs. 2), while on Wav2vec 2.0 producing WER reduction of {\bf 4.37\% absolute (16.53\% relative)} (Table~\ref{tab:recog-UASpeech-W2V}, Sys. 3 vs. 2).

\vspace{-0.25em}
\subsection{System Combination}
The best-performing hybrid DNN, E2E Conformer and fine-tuned Wav2vec 2.0 systems are further combined using two-pass rescoring~\cite{cui2022two}, where the DNN system produces N-best outputs ($N=50$) in the $1^{st}$ decoding pass while the E2E systems perform $2^{nd}$ pass rescoring using score interpolation. The final system (Table~\ref{tab:recog-UASpeech-combine} Sys.3) gives an overall WER of 17.82\% (51.25\% on very low intelligibility) on the UASpeech test set of $16$ dysarthric speakers. To the best of our knowledge, this is the best performance reported so far on UASpeech.

\vspace{-0.8em}
\begin{table}[ht]
    \caption{Performance of combining the best DNN (Table~\ref{tab:recog-UASpeech-hybrid-2} Sys.14), E2E Conformer (Table~\ref{tab:recog-UASpeech-Conformer} Sys.3) and finetuned Wav2vec 2.0 (Table~\ref{tab:recog-UASpeech-W2V} Sys.3) systems via two-pass rescoring~\cite{cui2022two}.}
    \label{tab:recog-UASpeech-combine}
    \Large
    \centering
    \vspace{-0.5em}
    \setlength{\abovecaptionskip}{0.05cm}
    \renewcommand\arraystretch{1.0}
    \renewcommand\tabcolsep{2.0pt}
    \scalebox{0.45}{\begin{tabular}{c|c|c|cccc|cc|c}
    \hline\hline  
        \multirow{3}{*}{Sys.} & 
        \multirow{3}{*}{Model} & 
        \multirow{3}{*}{\#Hrs} &
        \multicolumn{7}{c}{WER\%} \\
    \cline{4-10}
        & & &
        \multicolumn{4}{c|}{Intel. Subgroup} &
        \multirow{2}{*}{Unseen} & 
        \multirow{2}{*}{Seen} &
        \multirow{2}{*}{All} \\ 
    \cline{4-7}
        & & & VL & L & M & H & & & \\
    \hline\hline
        1 & DNN $\rightarrow$ CONF. & \multirow{3}{*}{190} & 60.66 & 22.82 & 11.37 & 3.80 & 31.62 & 16.15 & 22.22 \\
    \cline{1-2}\cline{4-10}
        2 & DNN $\rightarrow$ W2V. & & 52.00 & 18.26 & 8.35 & 2.58 & 26.41 & 12.92 & 18.21 \\
    \cline{1-2}\cline{4-10}
        3 & DNN $\rightarrow$ CONF. $+$ W2V. & & \textbf{51.25} & \textbf{17.41} & \textbf{8.16} & \textbf{2.66} & \textbf{26.45} & \textbf{12.26} & \textbf{17.82} \\
    \hline\hline 
    \end{tabular}}
\end{table}

\vspace{-1.5em}
\begin{table}[!h]
  \caption{Performance comparison against recently published systems on \textbf{UASpeech}. ``DA'' denotes data augmentation.}
  \label{tab:compare}
  \Large
  \centering
  \vspace{-0.5em}
  \renewcommand\arraystretch{1}
  \scalebox{0.42}{\begin{tabular}{ccc}
  \toprule
    Sys. & VL & All \\
  \midrule
    Sheffield-2020 Fine-tuning CNN-TDNN speaker adaptation (15spk)~\cite{xiong2020source} & 68.24 & 30.76 \\
    CUHK-2020 DNN + DA + LHUC SAT~\cite{geng2020investigation} & 62.44 & 26.37 \\ 
    CUHK-2021 QuartzNet + CTC + Meta-learning + SAT~\cite{wang2021improved} & 69.30 & 30.50 \\
    Sheffield-2022 DA + Source Filter Features + iVector adapt~\cite{yue2022acoustic} & - & 30.30 \\
    BUT-2022 Wav2vec 2.0 + fMLLR + xVectors (15spk)~\cite{baskar2022speaker} & 57.72 & 22.48 \\
    CUHK-2022 DA + TDNN + Wav2vec 2.0 feat. + system combination~\cite{hu2022exploring} & 52.53 & 22.83 \\
    \textbf{DA + severity incorporation + system combination (Table~\ref{tab:recog-UASpeech-combine} Sys.3, ours)} & \textbf{51.25} & \textbf{17.82} \\
  \bottomrule
  \end{tabular}}
  \vspace{-4mm}
\end{table}

\vspace{-0.5em}
\section{Conclusions}
\label{sec:conc}
This paper investigates a novel set of techniques to incorporate speech impairment severity into state-of-the-art hybrid DNN, E2E Conformer and pre-trained Wav2vec 2.0 systems for dysarthric speech recognition. These techniques include the use of multitask learning, auxiliary features and structured LHUC transforms. Experiments conducted on the UASpeech dataset suggest that incorporating severity and combining the best-performing systems produces the lowest published WER of 17.82\%. Future research will focus on more advanced techniques of incorporating severity into the E2E systems.

% \vspace{-0.5em}
% \section{Acknowledgements}
% This research is supported by Hong Kong RGC GRF grant No. 14200021, 14200218, 14200220, TRS T45-407/19N and Innovation \& Technology Fund grant No. ITS/254/19 and National Natural Science Foundation of China (NSFC) Grant 62106255.

\bibliographystyle{IEEEtran}
\bibliography{main}

\begin{thebibliography}{10}
\providecommand{\url}[1]{#1}
\csname url@samestyle\endcsname
\providecommand{\newblock}{\relax}
\providecommand{\bibinfo}[2]{#2}
\providecommand{\BIBentrySTDinterwordspacing}{\spaceskip=0pt\relax}
\providecommand{\BIBentryALTinterwordstretchfactor}{4}
\providecommand{\BIBentryALTinterwordspacing}{\spaceskip=\fontdimen2\font plus
\BIBentryALTinterwordstretchfactor\fontdimen3\font minus
  \fontdimen4\font\relax}
\providecommand{\BIBforeignlanguage}[2]{{%
\expandafter\ifx\csname l@#1\endcsname\relax
\typeout{** WARNING: IEEEtran.bst: No hyphenation pattern has been}%
\typeout{** loaded for the language `#1'. Using the pattern for}%
\typeout{** the default language instead.}%
\else
\language=\csname l@#1\endcsname
\fi
#2}}
\providecommand{\BIBdecl}{\relax}
\BIBdecl

\bibitem{christensen2012comparative}
H.~Christensen \emph{et~al.}, ``{A comparative study of adaptive, automatic
  recognition of disordered speech},'' in \emph{INTERSPEECH}, 2012.

\bibitem{kim2013dysarthric}
M.~J. Kim \emph{et~al.}, ``{Dysarthric speech recognition using
  dysarthria-severity-dependent and speaker-adaptive models},'' in
  \emph{INTERSPEECH}, 2013.

\bibitem{sehgal2015model}
S.~Sehgal \emph{et~al.}, ``{Model adaptation and adaptive training for the
  recognition of dysarthric speech},'' in \emph{SLPAT}, 2015.

\bibitem{joy2018improving}
N.~M. Joy \emph{et~al.}, ``{Improving acoustic models in TORGO dysarthric
  speech database},'' \emph{IEEE T NEUR SYS REH}, 2018.

\bibitem{yu2018development}
J.~Yu \emph{et~al.}, ``{Development of the CUHK Dysarthric Speech Recognition
  System for the UASpeech Corpus},'' in \emph{INTERSPEECH}, 2018.

\bibitem{hu2019cuhk}
S.~Hu \emph{et~al.}, ``{The CUHK Dysarthric Speech Recognition Systems for
  English and Cantonese},'' in \emph{INTERSPEECH}, 2019.

\bibitem{xiong2019phonetic}
F.~Xiong \emph{et~al.}, ``{Phonetic analysis of dysarthric speech tempo and
  applications to robust personalised dysarthric speech recognition},'' in
  \emph{ICASSP}, 2019.

\bibitem{liu2020exploiting}
S.~Liu \emph{et~al.}, ``{Exploiting cross-domain visual feature generation for
  disordered speech recognition},'' in \emph{INTERSPEECH}, 2020.

\bibitem{geng2020investigation}
M.~Geng \emph{et~al.}, ``{Investigation of Data Augmentation Techniques for
  Disordered Speech Recognition},'' in \emph{INTERSPEECH}, 2020.

\bibitem{harvill2021synthesis}
J.~Harvill \emph{et~al.}, ``{Synthesis of new words for improved dysarthric
  speech recognition on an expanded vocabulary},'' in \emph{ICASSP}, 2021.

\bibitem{geng2021spectro}
M.~Geng \emph{et~al.}, ``{Spectro-Temporal Deep Features for Disordered Speech
  Assessment and Recognition},'' in \emph{INTERSPEECH}, 2021.

\bibitem{xie2021variational}
X.~Xie \emph{et~al.}, ``{Variational Auto-Encoder Based Variability Encoding
  for Dysarthric Speech Recognition},'' in \emph{INTERSPEECH}, 2021.

\bibitem{liu2021recent}
S.~Liu \emph{et~al.}, ``{Recent Progress in the CUHK Dysarthric Speech
  Recognition System},'' \emph{IEEE T AUDIO SPEECH}, 2021.

\bibitem{jin2021adversarial}
Z.~Jin \emph{et~al.}, ``{Adversarial Data Augmentation for Disordered Speech
  Recognition},'' in \emph{INTERSPEECH}, 2021.

\bibitem{deng2021bayesian}
J.~Deng \emph{et~al.}, ``{Bayesian Parametric and Architectural Domain
  Adaptation of LF-MMI Trained TDNNs for Elderly and Dysarthric Speech
  Recognition},'' in \emph{INTERSPEECH}, 2021.

\bibitem{hu2021neural}
S.~Hu \emph{et~al.}, ``{Neural Architecture Search For LF-MMI Trained Time
  Delay Neural Networks},'' \emph{IEEE T AUDIO SPEECH}, 2022.

\bibitem{geng2022fly}
M.~Geng \emph{et~al.}, ``On-the-fly feature based speaker adaptation for
  dysarthric and elderly speech recognition,'' \emph{arXiv preprint
  arXiv:2203.14593}, 2022.

\bibitem{jin2022personalized}
Z.~Jin \emph{et~al.}, ``Personalized adversarial data augmentation for
  dysarthric and elderly speech recognition,'' \emph{arXiv preprint
  arXiv:2205.06445}, 2022.

\bibitem{mulfari2022exploring}
D.~Mulfari \emph{et~al.}, ``Exploring ai-based speaker dependent methods in
  dysarthric speech recognition,'' in \emph{CCGrid}, 2022.

\bibitem{hu2022exploiting}
S.~Hu \emph{et~al.}, ``Exploiting cross-domain and cross-lingual ultrasound
  tongue imaging features for elderly and dysarthric speech recognition,''
  \emph{arXiv preprint arXiv:2206.07327}, 2022.

\bibitem{hu2022exploit}
S.~Hu \emph{et~al.}, ``{Exploiting Cross Domain Acoustic-to-articulatory
  Inverted Features for Disordered Speech Recognition},'' in \emph{ICASSP},
  2022.

\bibitem{tobin2022personalized}
J.~Tobin \emph{et~al.}, ``Personalized automatic speech recognition trained on
  small disordered speech datasets,'' in \emph{ICASSP}, 2022.

\bibitem{baskar2022speaker}
M.~K. Baskar \emph{et~al.}, ``{Speaker adaptation for Wav2vec2 based dysarthric
  ASR},'' in \emph{INTERSPEECH}, 2022.

\bibitem{hernandez2022cross}
A.~Hernandez \emph{et~al.}, ``Cross-lingual self-supervised speech
  representations for improved dysarthric speech recognition,'' in
  \emph{INTERSPEECH}, 2022.

\bibitem{geng2022spectro}
M.~Geng \emph{et~al.}, ``{Speaker Adaptation Using Spectro-Temporal Deep
  Features for Dysarthric and Elderly Speech Recognition},'' \emph{IEEE T AUDIO
  SPEECH}, 2022.

\bibitem{yue2022acoustic}
Z.~Yue \emph{et~al.}, ``Acoustic modelling from raw source and filter
  components for dysarthric speech recognition,'' \emph{IEEE T AUDIO SPEECH},
  2022.

\bibitem{jin2022adversarial}
Z.~Jin \emph{et~al.}, ``Adversarial data augmentation using vae-gan for
  disordered speech recognition,'' in \emph{ICASSP}, 2023.

\bibitem{hu2022exploring}
S.~Hu \emph{et~al.}, ``Exploring self-supervised pre-trained asr models for
  dysarthric and elderly speech recognition,'' in \emph{ICASSP}, 2023.

\bibitem{kent2000dysarthrias}
R.~D. Kent \emph{et~al.}, ``{What dysarthrias can tell us about the neural
  control of speech},'' \emph{J PHONETICS}, 2000.

\bibitem{kodrasi2020spectro}
I.~Kodrasi \emph{et~al.}, ``{Spectro-Temporal Sparsity Characterization for
  Dysarthric Speech Detection},'' \emph{IEEE T AUDIO SPEECH}, 2020.

\bibitem{bhat2016recognition}
C.~Bhat \emph{et~al.}, ``{Recognition of Dysarthric Speech Using Voice
  Parameters for Speaker Adaptation and Multi-Taper Spectral Estimation},'' in
  \emph{INTERSPEECH}, 2016.

\bibitem{kim2017regularized}
M.~Kim \emph{et~al.}, ``{Regularized speaker adaptation of KL-HMM for
  dysarthric speech recognition},'' \emph{IEEE T NEUR SYS REH}, 2017.

\bibitem{takashima2020two}
R.~Takashima \emph{et~al.}, ``{Two-step acoustic model adaptation for
  dysarthric speech recognition},'' in \emph{ICASSP}, 2020.

\bibitem{mustafa2014severity}
M.~B. Mustafa \emph{et~al.}, ``Severity-based adaptation with limited data for
  asr to aid dysarthric speakers,'' \emph{PloS one}, 2014.

\bibitem{bhat2016improving}
C.~Bhat \emph{et~al.}, ``Improving recognition of dysarthric speech using
  severity based tempo adaptation,'' in \emph{SPECOM 2016}, 2016.

\bibitem{soleymanpour2022synthesizing}
M.~Soleymanpour \emph{et~al.}, ``Synthesizing dysarthric speech using
  multi-speaker tts for dysarthric speech recognition,'' in \emph{ICASSP},
  2022.

\bibitem{gulati2020conformer}
A.~Gulati \emph{et~al.}, ``{Conformer: Convolution-augmented Transformer for
  Speech Recognition},'' in \emph{INTERSPEECH}, 2020.

\bibitem{baevski2020wav2vec}
A.~Baevski \emph{et~al.}, ``wav2vec 2.0: A framework for self-supervised
  learning of speech representations,'' \emph{ADV NEUR IN}, 2020.

\bibitem{caruana1997multitask}
R.~Caruana, ``{Multitask learning},'' \emph{MACH LEARN}, 1997.

\bibitem{swietojanski2016learning}
P.~Swietojanski \emph{et~al.}, ``{Learning hidden unit contributions for
  unsupervised acoustic model adaptation},'' \emph{IEEE T AUDIO SPEECH}, 2016.

\bibitem{kim2008dysarthric}
H.~Kim \emph{et~al.}, ``{Dysarthric speech database for universal access
  research},'' in \emph{INTERSPEECH}, 2008.

\bibitem{pallet1990tools}
D.~S. Pallet \emph{et~al.}, ``Tools for the analysis of benchmark speech
  recognition tests,'' in \emph{ICASSP}, 1990.

\bibitem{cui2022two}
M.~Cui \emph{et~al.}, ``Two-pass decoding and cross-adaptation based system
  combination of end-to-end conformer and hybrid tdnn asr systems,'' in
  \emph{INTERSPEECH}, 2022.

\bibitem{senior2014improving}
A.~Senior \emph{et~al.}, ``{Improving DNN speaker independence with i-vector
  inputs},'' in \emph{ICASSP}, 2014.

\bibitem{snyder2018x}
D.~Snyder \emph{et~al.}, ``{X-vectors: Robust dnn embeddings for speaker
  recognition},'' in \emph{ICASSP}, 2018.

\bibitem{pironkov2016multi}
G.~Pironkov \emph{et~al.}, ``Multi-task learning for speech recognition: an
  overview.'' in \emph{ESANN}, 2016.

\bibitem{yu2013kl}
D.~Yu \emph{et~al.}, ``Kl-divergence regularized deep neural network adaptation
  for improved large vocabulary speech recognition,'' in \emph{ICASSP}, 2013.

\bibitem{graves2006connectionist}
A.~Graves \emph{et~al.}, ``Connectionist temporal classification: labelling
  unsegmented sequence data with recurrent neural networks,'' in \emph{ICML},
  2006.

\bibitem{vaswani2017attention}
A.~Vaswani \emph{et~al.}, ``Attention is all you need,'' \emph{ADV NEUR IN},
  2017.

\bibitem{young2002htk}
S.~Young \emph{et~al.}, ``{The HTK book},'' \emph{Cambridge university
  engineering department}, 2002.

\bibitem{povey2011kaldi}
D.~Povey \emph{et~al.}, ``{The Kaldi speech recognition toolkit},'' in
  \emph{ASRU}, 2011.

\bibitem{watanabe2018espnet}
S.~Watanabe \emph{et~al.}, ``{ESPnet: End-to-End Speech Processing Toolkit},''
  in \emph{INTERPSEECH}, 2018.

\bibitem{xiong2020source}
F.~Xiong \emph{et~al.}, ``{Source Domain Data Selection for Improved Transfer
  Learning Targeting Dysarthric Speech Recognition},'' in \emph{ICASSP}, 2020.

\bibitem{wang2021improved}
D.~Wang \emph{et~al.}, ``{Improved end-to-end dysarthric speech recognition via
  meta-learning based model re-initialization},'' in \emph{ISCSLP}, 2021.

\end{thebibliography}

\end{document}